\documentclass[conference]{IEEEtran}
\renewcommand\IEEEkeywordsname{Keywords}

\usepackage{tikz}
\usetikzlibrary{arrows.meta,
                chains,
                positioning,
                shapes}
\tikzset {
    roundedbox/.style={rounded rectangle,draw=black,align=center},
}

\tikzstyle{terminator} = [rectangle, draw, text centered, rounded corners, minimum height=2em]
\tikzstyle{process} = [rectangle, draw, text centered, minimum height=2em]
\tikzstyle{decision} = [diamond, draw, text centered, minimum height=2em]
\tikzstyle{data}=[trapezium, draw, text centered, trapezium left angle=60, trapezium right angle=120, minimum height=2em]
\tikzstyle{connector} = [draw, -latex']
\usetikzlibrary{matrix}
\usepackage{tikz-qtree}
\usepackage{forest}
\usepackage{adjustbox}
\usepackage{float}
\usepackage[nocolor]{drawstack}
\usepackage{listings}
\usepackage{syntax}

\usepackage{stackengine,listofitems}
\def\wboxwidth{.7em}
\def\wboxstrut{\vphantom{w}}
\newcommand\wbox[1]{\fbox{\makebox[\wboxwidth]{#1\wboxstrut}}}
\newcommand\wstack[1]{%
  \setsepchar{,}%
  \setstackEOL{,}%
  \savestack\tmp{\Shortstack{#1}}%
  \def\wboxwidth{\wd\tmpcontent}%
  \readlist\boxitems{#1}%
  \savestack\boxbuild{\wbox{\boxitems[-1]}}%
  \foreachitem\x\in\boxitems{%
    \ifnum\xcnt=1\relax\else%
      \savestack\boxbuild{\stackon[-\fboxrule]{\boxbuild}{\wbox{\boxitems[-\xcnt]}}}%
    \fi%
  }%
  \boxbuild%
}

\newcommand{\code}[1]{\texttt{#1}}
\newcommand{\codebf}[1]{\textbf{\texttt{#1}}}

\newcounter{nodeidx}
\newcommand{\nodes}[1]{%
    \foreach \num in {#1}{
      \node[minimum size=6mm, rectangle] (\arabic{nodeidx}) at (\arabic{nodeidx},0) {\num};
      \stepcounter{nodeidx}
    }
}

\newcommand{\boxnodes}[1]{%
    \foreach \num in {#1}{
      \node[minimum size=6mm, draw, rectangle] (\arabic{nodeidx}) at (\arabic{nodeidx},0) {\num};
      \stepcounter{nodeidx}
    }
}

\newcommand{\brckt}[4]{
  \draw (#1.south west) ++($(-.1, -.1) + (-.1*#3, 0)$) -- ++($(0,-.1) + (0,-#3*1.25em)$) -- node [below] {#4} ($(#2.south east) + (.1,-.1) + (.1*#3, 0) + (0,-.1) + (0,-#3*1.25em)$) -- ++($(0,#3*1.25em) + (0,.1)$);%
}

\usepackage{amsthm}
\newtheorem{definition}{Definition}[section]

\usepackage{filecontents,lipsum}
\usepackage[noadjust]{cite}
\usepackage{amsmath}
\usepackage{amssymb}
\newcommand{\mat}[1]{\mathbf{#1}}
\ifCLASSOPTIONcompsoc
 \usepackage[caption=false,font=normalsize,labelfont=sf,textfont=sf]{subfig}
\else
 \usepackage[caption=false,font=footnotesize]{subfig}
\fi
\usepackage{flushend}
\usepackage{tabularx,booktabs,textcomp}
\newcolumntype{C}{>{\centering\arraybackslash}X} 
\newcolumntype{L}{>{\raggedright\arraybackslash}X} 
\usepackage[noadjust]{cite}

\usepackage{wasysym}
\usepackage{lipsum} 

\begin{document}

\title{Tensor Algebra on an Optoelectronic Microchip}

\author{
    \IEEEauthorblockN{Sathvik Redrouthu}
    \IEEEauthorblockA{Procyon Photonics\\
    Ashburn, Virginia\\
    Email: 2024sredrout@tjhsst.edu}
    \and
    \IEEEauthorblockN{Rishi Athavale}
    \IEEEauthorblockA{Procyon Photonics\\
    Ashburn, Virginia\\
    Email: rishi.athavale1@gmail.com}
}

\maketitle

\begin{abstract}

Tensor algebra lies at the core of computational science and machine learning. Due to its high usage, entire libraries exist dedicated to improving its performance. Conventional tensor algebra performance boosts focus on algorithmic optimizations, which in turn lead to incremental improvements.  In this paper, we describe a method to accelerate tensor algebra a different way: by outsourcing operations to an optical microchip. We outline a numerical programming language developed to perform tensor algebra computations that is designed to leverage our optical hardware's full potential. We introduce the language's current grammar and go over the compiler design. We then show a new way to store sparse rank-\textit{n} tensors in RAM that outperforms conventional array storage (used by C++, Java, etc.). This method is more memory-efficient than Compressed Sparse Fiber (CSF) format and is specifically tuned for our optical hardware. Finally, we show how the scalar-tensor product, rank-$n$ Kronecker product, tensor dot product, Khatri-Rao product, face-splitting product, and vector cross product can be compiled into operations native to our optical microchip through various tensor decompositions.

\end{abstract}

\section{Introduction}
\label{sec:intro}

\subsection{Tensor Algebra}
\label{subsec:talgintro}

Tensor algebra has numerous applications in scientific disciplines. For example, widely used multiphysics simulation software (e.g. COMSOL Multiphysics, Ansys Lumerical, etc.) must perform large-scale numerical computations to solve problems in numerous fields such as fluid dynamics, structural mechanics, heat transfer and electromagnetics \cite{comsol, ansys, multiphysics-modeling-veryst-engineering}. Many of these computations are streamlined through chained tensor algebra expressions \cite{MultiphysicsSimulation}. In addition, advances in machine learning (ML) due to large neural networks (e.g., DALL-E 2, GPT-3, PaLM, etc.) also make use of massive tensor algebra computations \cite{pmlr-v139-blalock21a}. Optimizing tensor algebra becomes exceedingly important when ML models must meet time constraints (e.g., high-frequency stock trading bots) \cite{DBLP:journals/corr/abs-2101-07107}.

Tensors themselves can be thought of as \textit{n}-dimensional numerical arrays for the purposes of this paper. Each dimension of a tensor is referred to as a mode. A tensor's rank is the number of modes it has and therefore the number indices needed to access a specific value \cite{parker:2016:meng-thesis}. Rank-0 tensors, having 0 modes, require no indices to access values and thus represent a single number, or a scalar. Similarly, rank-1 tensors are simply vectors and rank-2 tensors are matrices.

Tensors of rank $n>0$ are very useful in representing indexed data. For example, a search engine tracking page URLs, keywords, and backlinks can store collected data in a rank-3 tensor. Typically, however, not every element in this tensor is useful. It is often not the case that any given website contains each keyword and backlink ever indexed by the search engine. In the frequent scenario where a page URL does not map to a specific keyword-backlink combination, a 0 can simply be placed at \code{tensor[URL][keyword][backlink]}. This results in most of the tensor's entries becoming 0; such a tensor is referred to as a \textit{sparse} tensor \cite{NEURIPS2021-b0ab42fc}. We discuss efficient storage methods for sparse tensors in Sec. \ref{sec:tstor}.

Of course, search giants such as Google collect much more information than described in the example. Other companies are in the same boat; in fact, according to \cite{kjolstad:2017:taco}, a specific rank-3 Facebook tensor has dimensions $1591 \times 63891 \times 63890$. Huge computations are performed constantly on tensors like these; such is the case for most large-scale graph applications \cite{doi:10.1137/1.9780898719918.ch7}. Even after numerous algorithmic optimizations, however, such computation is far too slow to keep up with increasing demands \cite{https://doi.org/10.7298/5ksm-sm92}. For example, animation firms like Pixar can take up to 39 hours of computing time to render a single frame \cite{lehrer-2010}. It is therefore apparent that some form of optimization sustainable throughout the future is necessary.


\begin{figure}
    \centering
    \begin{tikzpicture}[auto matrix/.style={matrix of nodes,
        draw,thick,inner sep=0pt,
        nodes in empty cells,column sep=-0.2pt,row sep=-0.2pt,
        cells={nodes={minimum width=1.9em,minimum height=1.9em,
        draw,very thin,anchor=center,fill=white,
        execute at begin node={}
        }}}]
        \matrix[auto matrix=z,xshift=3em,yshift=3em](matz){
        3.1 & 0 & 0 & 0.9 & 0 \\
        0 & 0 & 2.3 & 0 & 1.3 \\
        0 & 0 & 0 & 0 & 0 \\
        };
        \matrix[auto matrix=y,xshift=1.5em,yshift=1.5em](maty){
        0.1 & 0 & 0 & 0 & 0 \\
        8.0 & 9.9 & 4.4 & 0 & 0 \\
        0 & 0 & 0 & 0 & 0 \\
        };

        \draw[thick,-stealth] ([xshift=1ex]maty.south east) -- ([xshift=1ex]matz.south east)
         node[midway,below] {$z$};
        \draw[thick,-stealth] ([yshift=-1ex]maty.south west) -- 
         ([yshift=-1ex]maty.south east) node[midway,below] {$x$};
        \draw[thick,-stealth] ([xshift=-1ex]maty.north west)
           -- ([xshift=-1ex]maty.south west) node[midway,above,rotate=90] {$y$};
    \end{tikzpicture}

    \caption{Example of a $5 \times 3 \times 2$ tensor. The tensor is referred to as "sparse" as most of the entries are 0. Most tensors encountered are sparse.}
    \label{fig:tens1}
\end{figure}
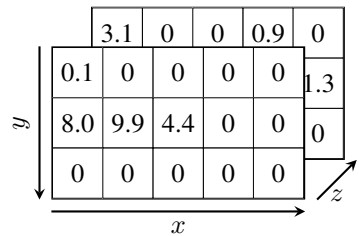



\subsection{The Photonic Advantage}

Many highly optimized tensor algebra libraries currently exist (e.g., Eigen, MATLAB Tensor Toolbox, and SPLATT) \cite{DBLP:journals/corr/abs-1911-12604, osti-1349514, SPLATT}. However, as Moore's Law and Dennard Scaling reach their limits and the demand for tensor algebra increases, running tensor algebra on classical hardware will no longer be viable and these libraries must adapt \cite{https://doi.org/10.7298/5ksm-sm92}.

An alternative to classical hardware involves optical computing (the use of photons to perform computations), which offers a significant speed increase and surmounts most of the energy challenges posed in conventional computer engineering \cite{Cole:21}. Moreover, its lack of dependence on the conventional transistor leads it to be independent from the decline of Moore's Law. Recognizing this, some of us at Procyon Photonics have designed an optical microchip able to perform high-speed matrix-vector multiplication (MVM). The chip (named Tachyon 1) maintains a compact form and is inherently analog, indicating its potential in computational fields \cite{DBLP:journals/corr/abs-2102-06365}.

Performing tensor algebra on such a microchip would offer a significant speed increase while simultaneously sidestepping the decline of Moore's Law. In this paper, we describe a method where this is possible.

\subsection{Apollo}

To our knowledge, no programming language has been invented that can leverage an optical microchip’s full potential and link it to fields that can be influenced by its capabilities. For these reasons, we introduce Apollo, a computing language designed specifically for Tachyon 1. Apollo supports important tensor algebra operations that are mapped onto the corresponding units on the host computer and optical chip. The language will be extended to support operations and algorithms that are not related solely to tensor algebra but still important for computationally expensive tasks, such as deep neural network (DNN) training/inference.

We begin by going through preliminary notation and definitions in Sec \ref{sec:prelim}. Next, we cover the language's grammar and supported operations in Sec. \ref{sec:ldet}. In Sec. \ref{sec:cdes}, we go over the workflow, compiler front-end, and virtual machine (VM). It is here where we introduce the most important VM instruction that Sec. \ref{sec:decomp} revolves around.

Next, we illustrate a new method to store large, sparse tensors in Sec. \ref{sec:tstor}, which we found to surpass the conventional array storage method from a memory viewpoint. In addition, we show how our method is more efficient than CSF format for our optical hardware. Finally, since Tachyon 1 is engineered to perform matrix-vector multiplication in a single instruction, we focus on decomposing complex tensor algebra expressions into sequences of matrix-vector products in Sec. \ref{sec:decomp}. Efficient tensor decompositions would allow entire tensor algebra expressions to be run at an incredible speed.

\section{Preliminaries}
\label{sec:prelim}
\subsection{Notation}

Tensors of rank $n>2$ are denoted in scripted letters (e.g., $\mathcal{X}$). Matrices are denoted in uppercase boldface and vectors are denoted in lowercase boldface ($\mat{M}$ and $\mat{v}$ respectively). The identity matrix is denoted as $\mat{I}$.

\subsection{Definitions}
We use multiple tensor operations in Apollo, some of which are modifications of existing definitions. In this section, we define each operation the way it is used within the language.

\begin{definition}[Scalar-tensor product]
\label{def:stprod}
Given a scalar $\lambda$ and a tensor $\mathcal{X} \in \mathbb{R}^{I_1 \times I_2 \times \dots \times I_n}$, the scalar-tensor product $\lambda\mathcal{X} \in \mathbb{R}^{I_1 \times I_2 \times \dots \times I_n}$ is given by:
\[
(\lambda\mathcal{X})_{i_1i_2 \dots i_n} = \lambda(x_{i_1i_2 \dots i_n})
\]
\end{definition}
\begin{definition}[Rank-$n$ Kronecker product]
\label{def:kron}
Given two tensors $\mathcal{X} \in \mathbb{R}^{I_1 \times I_2 \times \dots \times I_n}$ and $\mathcal{Y} \in \mathbb{R}^{J_1 \times J_2 \times \dots \times J_n}$, the rank-$n$ Kronecker product $\mathcal{X} \otimes \mathcal{Y} \in \mathbb{R}^{I_1J_1 \times I_2J_2 \times \dots I_nJ_n}$ is given by:
\[
(\mathcal{X} \otimes \mathcal{Y})_{i_1i_2 \dots i_n} = (x_{j_1j_2 \dots j_n})\mathcal{Y}
\]
Each index $i_1i_2 \dots i_n$ is a corresponding index in a block tensor. 
\end{definition}
\begin{definition}[Tensor inner product]
\label{def:tinner}
Given two tensors $\mathcal{X}, \mathcal{Y} \in \mathbb{R}^{I_1 \times I_2 \times \dots \times I_n}$, the inner product $\langle \mathcal{X} \;, \; \mathcal{Y} \rangle \in \mathbb{R}$ is given by:
\[
\langle \mathcal{X} \;, \; \mathcal{Y} \rangle = \sum_{i_1} \sum_{i_2} \cdots \sum_{i_n} x_{i_1i_2 \dots i_n} y_{i_1i_2 \dots i_n}
\]
\end{definition}
\begin{definition}[Tensor dot product]
\label{def:tdot}
Given two tensors $\mathcal{X} \in \mathbb{R}^{I_1 \times I_2 \dots \times I_m}$ and $\mathcal{Y} \in \mathbb{R}^{J_1 \times J_2 \times \dots \times J_{n-1} \times J_n}$, the tensor dot product $\mathcal{X} \cdot \mathcal{Y} \in \mathbb{R}^{I_1 \times I_2 \times \dots \times I_{m-1} \times J_1 \times J_2 \times \dots \times J_{n-2} \times J_n}$ is given by:
\[
(\mathcal{X} \cdot \mathcal{Y})_{i_1i_2 \dots i_mj_1j_2 \dots j_n} = \sum_{i_m, j_{n-1}} x_{i_1i_2 \dots i_m} y_{j_1j_2 \dots j_{n-1}j_n}
\]
where $I_m=J_{n-1}$.
\end{definition}
\begin{definition}[Khatri-Rao product]
\label{def:krao}
Given two matrices $\mat{A} \in \mathbb{R}^{I \times K}$ and $\mat{B} \in \mathbb{R}^{J \times K}$, the Khatri-Rao product $\mat{A} \odot \mat{B} \in \mathbb{R}^{I \cdot J \times K}$ is given by:
\[
\mat{A} \odot \mat{B}=\begin{bmatrix}\mat{a_1} \otimes \mat{b_1} & \mat{a_2} \otimes \mat{b_2} & \cdots & \mat{a_K} \otimes \mat{b_K}\end{bmatrix}
\]
This can be thought of as a column-wise Kronecker product.
\end{definition}
\begin{definition}[Face-splitting product]
\label{def:fsplit}
Given two matrices $\mat{A} \in \mathbb{R}^{K \times I}$ and $\mat{B} \in \mathbb{R}^{K \times J}$, the face-splitting product $\mat{A} \bullet \mat{B} \in \mathbb{R}^{K \times I \cdot J}$ is given by:
\[
\mat{A} \bullet \mat{B}=
\begin{bmatrix}
\mat{a_1} \otimes \mat{b_1}\\
\mat{a_2} \otimes \mat{b_2}\\
\vdots\\
\mat{a_K} \otimes \mat{b_K}\\
\end{bmatrix}
\]
This can be thought of as a row-wise Kronecker product.
\end{definition}
\begin{definition}[Vector cross product]
\label{def:cross}
Given two vectors $\mat{u} \in \mathbb{R}^{3}$ and $\mat{v} \in \mathbb{R}^{3}$, the vector cross product $\mat{u} \times \mat{v} \in \mathbb{R}^{3}$ is given by:
\[
\mat{u} \times \mat{v} =
\begin{vmatrix}
\mat{e_1} & \mat{e_2} & \mat{e_3}\\
a_1 & a_2 & a_3\\
b_1 & b_2 & b_3\\
\end{vmatrix}
\]
\end{definition}

We discuss how to run each of these operations on our optical hardware in Sec. \ref{sec:decomp}.\footnote{We refer to Def. \ref{def:tinner} in the general case to provide a complete definition, but only discuss implementation in the vector case.}

\section{Language Details}
\label{sec:ldet}

\subsection{Grammar}
\label{subsec:grammar}

The language ideally would have a grammar that is intuitive, vast, and requires minimal coding on the user’s side. Since this is a prototype, however, the grammar is limited and technical. This minimizes the number of compiler tricks needed, which we found was a good avenue to take to focus on compiling tensor algebra expressions. The current grammar is described in Fig. \ref{fig:ebnf} in Extended Backus-Naur Form (EBNF) notation.

\begin{figure}
    \setlength{\grammarindent}{5em}
    
    \begin{grammar}
    
    <lower> $\Rightarrow$ `a' | `b' | `c' | `d' | `e' | `f' | `g' | `h' | `i' | `j' | `k' | `l' | `m' | `n' | `o' | `p' | `q' | `r' | `s' | `t' | `u' | `v' | `w' | `x' | `y' | `z'
    
    <upper> $\Rightarrow$ `A' | `B' | `C' | `D' | `E' | `F' | `G' | `H' | `I' | `J' | `K' | `L' | `M' | `N' | `O' | `P' | `Q' | `R' | `S' | `T' | `U' | `V' | `W' | `X' | `Y' | `Z'
    
    <digit> $\Rightarrow$ `0' | `1' | `2' | `3' | `4' | `5' | `6' | `7' | `8' | `9'
    
    <character> $\Rightarrow$ <lower> | <upper>
    
    \end{grammar}
    
    \begin{grammar}
    
    <integer> $\Rightarrow$ [+|-]<digit>\{<digit>\}
    
    <floating-point> $\Rightarrow$ [<integer>] `.' \{<integer>\}-
    
    <tensor> $\Rightarrow$ `{' [<tensor>] \{`,' <tensor>\} `}'
    
    <identifier> $\Rightarrow$ <character> \{<character> | digit | `_'\}
    
    \end{grammar}
    
    \begin{grammar}
    
    <primary> $\Rightarrow$ <integer> | <floating-point> | <identifier> | <tensor> | [-]<term>
    
    <factor> $\Rightarrow$ `(' <expr> `)' | <primary>
    
    <term> $\Rightarrow$ <factor> \{(`*' | `/' | `@' | `\&' | `\%' | `#') <factor>\}
    
    <expr> $\Rightarrow$ <term> \{(`+' | `-') <term>\}
    
    \end{grammar}
    
    \begin{grammar}
    
    <type> $\Rightarrow$ `int' | `float' | `tensor'
    
    <statement> $\Rightarrow$ `let' <type> <identifier> `=' <expr> `;'
    
    <program> $\Rightarrow$ \{<statement>\}
    
    \end{grammar}
    
    \caption{Apollo's grammar shown in EBNF. The base case in the recursive tensor structure is a list of comma separated integers and/or floating point values.}
    \label{fig:ebnf}
\end{figure}

Notable emphasis is placed on expressions, as they are the focus for our application of optical computing. Currently, is only possible to declare new statements, as shown Fig. \ref{fig:ebnf}. However, all operations we discuss are able to be performed with solely this grammar, which we plan to expand in the future.

\subsection{Supported Operators}

The standard PEMDAS order is supported for scalars. For tensors of rank $n \geq 1$, the order of operations should be defined with parenthesis. We show the operators supported in this Apollo prototype in tables \ref{binoptable} and \ref{unoptable}.

\begin{table}
\centering
\caption{Binary Operators}
\label{binoptable}
\begin{tabularx}{3.3in}{C C C}
\toprule
        Name & Operator & Usage\\ 
\midrule
        Addition & +   & $s_1 + s_2$ \\
        Subtraction & -   & $s_1 - s_2$ \\
        Multiplication, dot product & *   & $s_1s_2$, $s_1 \cdot \mathcal{T}_1$, $\mathcal{T}_1 \cdot s_1$, $\mathcal{T}_2 \cdot \mathcal{T}_2$\\
        Division & /   & $x \div y$ \\
        Kronecker product & @   & $\mathcal{T}_1 \otimes \mathcal{T}_2$  \\
        Khatri-Rao product & \& & $\mathcal{T}_1 \odot \mathcal{T}_2$ \\
        Face-splitting product & \% & $\mathcal{T}_1 \bullet \mathcal{T}_2$ \\
        Cross product & \#  & $\mat{x} \times \mat{y}$ \\
\bottomrule
\end{tabularx}

\end{table}

\begin{table}
\centering
\caption{Unary Operators}
\label{unoptable}
\begin{tabularx}{3.3in}{C C C}
\toprule
        Name & Operator & Usage\\ 
\midrule
        Negation & -   & $-x$ \\
\bottomrule
\end{tabularx}

\end{table}

\section{Compiler Design}
\label{sec:cdes}

\subsection{Workflow}

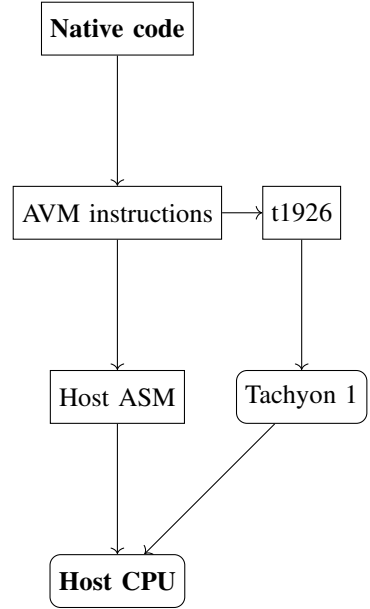
\begin{figure}[h!]
    \centering
    \begin{tikzpicture}[node distance = 2.45cm]
        \node [process] (native) {\textbf{Native code}};
        \node [process, below of=native] (ir) {AVM instructions};
        \node [process, right of=ir] (t1926) {t1926};
        \node [process, below of=ir] (x86) {Host ASM};
        \node [terminator, below of=t1926] (t1) {Tachyon 1};
        \node [terminator, below of=x86] (host) {\textbf{Host CPU}};
        \draw [->] (native) edge (ir) (ir) edge (t1926) (ir) edge (x86) (t1926) edge (t1) (x86) edge (host) (t1) edge (host);

    \end{tikzpicture}

    \caption{Native Apollo code gets compiled into Apollo Virtual Machine (AVM) instructions by the compiler front-end. AVM generates standard assembly instructions for regular operations and compiles tensor algebra to t1926 instructions. Respective assemblers target the host CPU and Tachyon 1. This chosen workflow enables tensor algebra to be outsourced to Tachyon 1. Note that the scope of this paper is limited to AVM instruction generation.}
    \label{fig:workflow}
\end{figure}

The workflow we decided on is shown in Fig. \ref{fig:workflow}. Note that the Apollo compiler is 2-stage.

The standard assembler targets the host CPU, whereas the t1926 assembler targets Tachyon 1. Such a setup is used because Tachyon 1 is geared towards certain types of computations only.

\subsection{Compiler Front-end}

We use a hand coded compiler front-end (lexer, parser, and code generator). This is because we have found that parser generators do not cooperate well with tensor algebra and our storage choice. We use a recursive descent parser, which works well for performance. The in-compiler tensor storage we describe in Sec. \ref{subsec:BSTT} is more easily implemented with such a parser.

It is noteworthy that there are many instances in the language where operators are overloaded. For example, consider the multiplication operator, \code{*}. If \code{A * B} is called, four cases are possible. 1) \code{A} is a scalar and \code{B} is a tensor of rank $n > 0$, 2) \code{A} is a tensor of rank $n > 0$ and \code{B} is a scalar, 3) \code{A} and \code{B} are both scalars, or 4) \code{A} and \code{B} are both tensors of rank $n > 0$. The parser considers these cases and generates abstract syntax tree (AST) nodes of the correct type (e.g., variable nodes, scalar nodes, tensor nodes, etc.).

The AST is traversed in pre-order by the code generator, sequentially producing the appropriate VM instructions. Standard procedures are followed for variable handling. In the case of more exotic AST nodes (e.g., tensor nodes) the code generator calls special functions (discussed in Sec. \ref{sec:decomp}) to generate the correct code. The VM instruction set is outlined in Sec. \ref{subsec:avm}.

\subsection{Virtual Machine}
\label{subsec:avm}

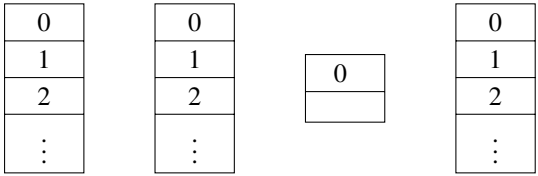
\begin{figure}[h!]
    \centering
    \begin{tikzpicture}[
        stack/.style={rectangle split, rectangle split parts=#1,draw, anchor=center}, align=center, node distance=2cm
    ]
    
    \node[stack=4, minimum width=3em] (A) {
        \nodepart{one}0
        \nodepart{two}1
        \nodepart{three}2
        \nodepart{four}\vdots
    };
    \node[stack=4, minimum width=3em, right of=A] (B) {
        \nodepart{one}0
        \nodepart{two}1
        \nodepart{three}2
        \nodepart{four}\vdots
    };
    \node[stack=2, minimum width=3em, right of=B] (C) {
        \nodepart{one}0
    };
    \node[stack=4, minimum width=3em, right of=C] (D) {
        \nodepart{one}0
        \nodepart{two}1
        \nodepart{three}2
        \nodepart{four}\vdots
    };
    
    \end{tikzpicture}%
    \caption{The \code{constant} (abstract), \code{global}, \code{pointer}, and \code{this} virtual segments respectively.}
    \label{fig:vsegments}
\end{figure}

Apollo’s VM is stack-based. It provides 4 memory segments (namely, the \code{constant}, \code{global}, \code{pointer}, and \code{this} segments), shown in Fig. \ref{fig:vsegments}\footnote{The RAM referred to throughout this section is a simplified virtual abstraction. Hence, we freely interact with it using numbers in the decimal system. The actual RAM is referred to when discussing compilation to target architectures, which will be done in a future paper.}.

Each one of these segments are anchored to a specific location in RAM at compile time\footnote{Exact RAM indices are not included.}.  They are fixed in their locations, except for the \code{this} segment, which we use for tensors. Index 0 in the \code{pointer} segment contains the base address of the \code{this} segment, so if the value at index 0 changes, the \code{this} segment gets anchored to a different RAM location, similar to \cite{nisan-schocken-2021}. As the language expands, we may add additional memory segments that can dynamically change location during run-time; if we take this route, we will allocate more RAM and add more values to the \code{pointer} segment.

The \code{constant} segment is used to push and pop constants to and from the stack, as in \cite{nisan-schocken-2021}. Note that despite showing solely integers, the segment supports integer and floating-point values. The \code{global} segment is used in conjunction with the symbol table to store variable values, which can be accessed throughout the lifetime of the program. Values in the \code{global} segment can also be references to tensors\footnote{Apollo does not yet support user-defined subroutines, so a \code{local} segment is not required.}. See Sec. \ref{sec:tstor} for more information regarding tensor storage.

The memory access commands are \codebf{push} \code{[segment] i} and  \codebf{pop} \code{[segment] i}. The push instruction pushes the value at index  \code{i} of memory segment  \code{[segment]} onto the stack. The pop instruction pops the value on top of the stack onto index  \code{i} of memory segment  \code{[segment]} \cite{nisan-schocken-2021}. The rest of the AVM instruction set (composed of arithmetic instructions and built-in subroutines) is given in Tables \ref{avmisa} and \ref{avmisasub}.

\begin{table}
\centering
\caption{AVM Arithmetic Instruction Set}
\label{avmisa}
\begin{tabularx}{3.3in}{C C C}
\toprule
        Operation & Compiles to & Description\\ 
\midrule
        \codebf{neg} & Host ASM & Negates the value at the top of stack. \\
        \codebf{add} & Host ASM & Pops stack into $b$. Pops stack into $a$. Pushes $a+b$ to stack. \\
        \codebf{sub} & Host ASM & Pops stack into $b$. Pops stack into $a$. Pushes $a-b$ to stack. \\
        \codebf{mult} & Host ASM & Pops stack into $b$. Pops stack into $a$. Pushes $ab$ to stack. \\
        \codebf{div} & Host ASM & Pops stack into $b$. Pops stack into $a$. Pushes $a \div b$ to stack.\\
        \codebf{mvmul} & Host ASM & Pops stack into $b$. Pops stack into $\mat{A}$. Pushes  $\mat{A}\mat{b}$  to stack. \\
\bottomrule
\end{tabularx}

\end{table}

\begin{table}
\centering
\caption{AVM Subroutine Instruction Set}
\label{avmisasub}
\begin{tabularx}{3.3in}{C C C}
\toprule
        Name & Args & Description\\ 
\midrule
        \code{malloc} & \codebf{int} \code{size} & Finds an unused RAM segment of length \code{size}, pushes pointer pointing to the first segment index to stack. \\
\bottomrule
\end{tabularx}

\end{table}

Note that each arithmetic instruction can be done in a single instruction by the corresponding processor.

Subroutines are handled with the instruction \codebf{call} \code{[fname] [nArgs]}. The first \code{[nArgs]} values are treated as arguments, so the virtual machine would pop the stack \code{[nArgs]} times if the call command is generated.

Since \code{malloc} has 1 argument, a possible code fragment for it looks like:

\begin{center}
\color{purple} \codebf{push} \color{black} \code{constant 3}\\
\color{purple} \codebf{call} \color{black} \code{malloc 1}
\end{center}

This would 1) push 3 onto the stack, 2) pop 3 off the stack and pass it into  \code{malloc}, 3) find an unused RAM segment of size 3, and 4) push a pointer to the first index of that segment to the RAM. Its behavior mimics  \code{Memory.alloc} in \cite{nisan-schocken-2021}.

\section{Sparse Tensor Storage}
\label{sec:tstor}

\subsection{Current Methods}

Tensor components are conventionally represented as nested arrays in standard programming languages. In C++, the components are stored as one contiguous array. To access the element at index $ij$, the element at index \code{base+i+j} in the flattened block is indexed (where \code{base} is the base address of the array) \cite{corob-msft}. In Java, each array of dimension $n+1$ contains pointers to each sub-array of dimension $n$. If $n=0$, the $(n+1)$-dimensional array simply stores scalar values \cite{java-array}.

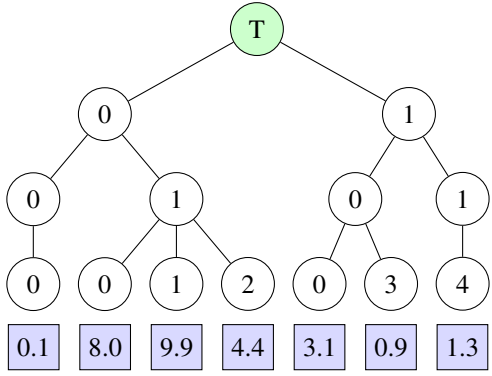
\begin{figure}
    \centering
    \begin{forest}
    for tree={circle, draw, minimum size=2em}
    [T, fill=green!20
        [0
            [0
                [0]
            ]
            [1
                [0]
                [1]
                [2]
            ]
        ]
        [1
            [0
                [0]
                [3]
            ]
            [1
                [4]
            ]
        ]
    ]
    \end{forest}
    \begin{adjustbox}{margin=5pt}
    \begin{tikzpicture}
        \coordinate (s) at (0,0);
        \foreach \num in {0.1,8.0,9.9,4.4,3.1,0.9,1.3}{
          \node[minimum size=6mm, draw, rectangle, fill=blue!15] at (s) {\num};
          \coordinate (s) at ($(s) + (0.95,0)$);
        }
    \end{tikzpicture}
    \end{adjustbox}
    
    \caption{CSF representation of the tensor in Fig. \ref{fig:tens1}, as proposed by \cite{SPLATT}.}
    \label{fig:csf}
    
\end{figure}

Since tensors are often sparse, however, these conventional methods often end up storing excess zeros, making them sub-optimal. The Facebook tensor discussed in Sec. \ref{subsec:talgintro} has only 737,934 nonzero values and is therefore 99.999\% made up of zeroes. It is apparent that tensor storage optimizations must be considered.  Compressed Sparse Fiber (CSF) format is a better method that stores a tensor in a tree structure, where the indices and values for only non-zero components are contained, as shown in Fig. \ref{fig:csf}. CSF performs significant better than conventional approaches for applications involved in highly sparse tensor algebra.

However, CSF requires storing pointers to each child node, likely integrated to enable fast indexing \cite{SPLATT}. Such an optimization would typically be incredibly important; however, since our optical hardware can do an MVM in a single instruction, it is not necessary that we are able to access indices efficiently in intermediate computations. Rather, it is important that we return an entire row of indices as fast as possible. Sec. \ref{sec:decomp} provides insight into why this is the case.

\subsection{Binary Sparse Tensor Tree Format}
\label{subsec:BSTT}

\begin{figure}
    \centering
    \begin{forest}
    for tree={circle, draw, minimum size=2em}
    [0, fill=green!20
        [0
            [0
                [0.1, fill=blue!15]
                [,phantom]
            ]
            [1
                [0
                    [8.0, fill=blue!15]
                    [1
                        [9.9, fill=blue!15]
                        [2
                            [4.4, fill=blue!15]
                            [,phantom]
                        ]
                    ]
                ]
                [,phantom]
            ]
        ]
        [1
            [0
                [0
                    [3.1, fill=blue!15]
                    [3
                        [0.9, fill=blue!15]
                        [,phantom]
                    ]
                ]
                [1
                    [4
                        [1.3, fill=blue!15]
                        [,phantom]
                    ]
                    [,phantom]
                ]
            ]
            [,phantom]
        ]
    ]
    \end{forest}
    \caption{Our representation of the tensor in Fig. \ref{fig:tens1}, which we refer to as Binary Sparse Tensor Tree (BSTT) format. Each non-leaf node contains an index. The left child is always the root of a sub-tensor belonging to the current tensor. The right child is always the root of the next sub-tensor belonging to the parent tensor shared by the current tensor.}
    \label{fig:bstt}
\end{figure}
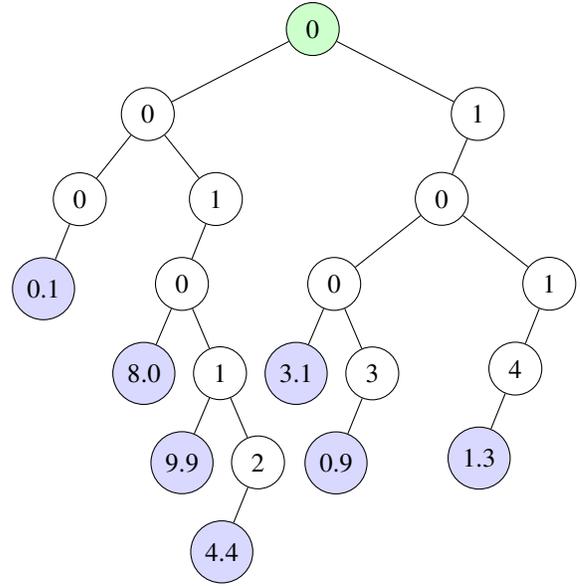

To save memory and return sub-tensors quickly, we store the tensor in Fig. \ref{fig:tens1} as shown in Fig. \ref{fig:bstt}.

We only use this format for intermediate computations. It is slower to index into a specific value, but this is irrelevant as such indexing is not necessary for Apollo-supported intermediate computations on Tachyon 1. Again, however, we must be able to access a full row of rank-$n$ indices easily. This is efficient with our format as we can simply return a pointer to the first index. Hence, our method is more useful than CSF format for our purposes.

\begin{figure}
    \begin{tikzpicture}
        \centering
        \pgftransparencygroup
        \boxnodes{0,0,0,0.1,1,0}
        \endpgftransparencygroup
        \pgftransparencygroup
        \nodes{$\cdots$}
        \endpgftransparencygroup
        \pgftransparencygroup
        \boxnodes{4,1.3}
        \endpgftransparencygroup
        \pgftransparencygroup
        \brckt{1}{9}{0}{size=20}
        \endpgftransparencygroup
        \pgftransparencygroup
        \endpgftransparencygroup
    \end{tikzpicture}

    \caption{Pre-order traversal of BSTT format results in the following array, which is then stored on the heap (we plan to make tensors mutable in future Apollo versions). Values are always assumed to be floating point numbers, a safe assumption due to the large number of non-integer values encountered in the targeted fields \cite{https://doi.org/10.7298/5ksm-sm92, parker:2016:meng-thesis, SPLATT, NEURIPS2021-b0ab42fc, kjolstad:2017:taco}. Indices are always integers. This allows us to determine the leaf nodes and "reconstruct" the tree when needed.}
    \label{fig:bsttstk}
\end{figure}
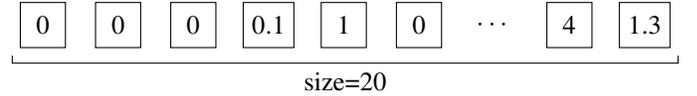

\section{Compiling Tensor Algebra Expressions}
\label{sec:decomp}

As stated in earlier sections, the most powerful tensor algebra operation supported by Tachyon 1 that can be done in a single instruction is matrix-vector multiplication (MVM). Therefore, it is the compiler’s job to translate more complex operations into sequences of MVMs when applicable, thereby accelerating computation of the whole expression. For clarity, note that Tachyon 1 multiplies matrices and vectors in the order $\mat{A}\mat{x}=\mat{b}$. Also note that decomposition into sub-tensors of who’s sizes are supported by Tachyon 1 is not covered in this paper.

\subsection{Scalar-tensor Product}
\label{subsec:stprod}

The scalar-tensor product as defined in Def. \ref{def:stprod} is a commutative operation that multiplies each element in a tensor $\mathcal{X}$ by a scalar $\lambda$\footnote{$\mathcal{X}$ is assumed to be a tensor of rank $n>0$, since the parser would map the scalar case to scalar multiplication.}. The product is very easy to compile; simply iterate through each vector $\mat{x}_{i_1i_2 \dots i_{n-1}}$ in the tensor and generate the \codebf{mvmul} instruction to multiply it by the matrix 
$\lambda \mat{I} = 
\begin{bmatrix}
\lambda & 0\\
0 & \lambda
\end{bmatrix}$.
Note that the compiler reorients the product to generate the matrix before the vector if the user calls it in the opposite order. In other words, it ensures that running the generated code results in a product in the order $\lambda \mat{I} \mat{x}_{i_1i_2 \dots i_{n-1}}$.

\subsection{Rank-$n$ Kronecker Product}
\label{subsec:kron}

The Kronecker product is useful in signal and image processing \cite{LOAN200085}. Through the Khatri-Rao product, it is useful in neural networks (through minimization of convolution and tensor sketch operations) and natural language processing \cite{DeepKNN}.

Refer to Def. \ref{def:kron} for the definition of the Kronecker product. For clarity, each element in the result is simply the element at $x_{i_1 i_2 \dots i_n}$ multiplied by $\mathcal{Y}$ for two tensors $\mathcal{X}$ and $\mathcal{Y}$. The product can be represented compactly between two matrices as

\[
\mat{A} \otimes \mat{B} = 
\begin{bmatrix}
a_{11}\mat{B} & a_{12}\mat{B} & \cdots & a_{1n}\mat{B}\\
a_{21}\mat{B} & a_{22}\mat{B} & \cdots & a_{2n}\mat{B}\\
\vdots & \vdots & \ddots & \vdots\\
a_{m1}\mat{B} & a_{m2}\mat{B} & \cdots & a_{mn}\mat{B}\\
\end{bmatrix}
\]
Therefore, the compiler can compute the scalar-tensor product for each element in the resultant block tensor through the method outlined in Section \ref{subsec:stprod}.

\subsection{Tensor Dot Product}
\label{subsec:tdot}

Many fields, including machine learning and physics, demand the ability to compute the dot product efficiently \cite{DotProductInML, DotProductInQuantumPhysics}. To allow Tachyon 1 to meet this demand, we must also provide a way for the Apollo compiler to transform this operation into a sequence of MVMs. The tensor dot product is an operation between two rank-$n$ tensors, $\mathcal{A}$ and $\mathcal{B}$. Some possibly familiar tensor dot products include the rank-0, rank-1, and rank-2 dot products (scalar product, vector dot product, and matrix multiplication respectively).
The compiler considers the dot product operation over the component arrays. A few cases are possible:
\begin{enumerate}
    \item $\mathcal{A}$ and $B$ are both scalars
    \item Either $\mathcal{A}$ or $\mathcal{B}$ is a scalar, but not both
    \item $\mathcal{A}$ is a vector/matrix, whereas $\mathcal{B}$ is a vector 
    \item $\mathcal{A}$ is a vector/matrix, whereas $\mathcal{B}$ is a rank-$n$ tensor with $n>2$
    \item $\mathcal{A}$ is a rank-$n$ tensor where $n>2$, whereas $\mathcal{B}$ is a vector
    \item $\mathcal{A}$ is a rank-$n$ tensor and $\mathcal{B}$ is a rank-$m$ tensor, where $n>1$, $m>1$, and $n\neq m$
    \item $\mathcal{A}$ and $\mathcal{B}$ are both rank-$n$ tensors
\end{enumerate}

Cases 1 and 2 are irrelevant since the parser maps Case 1 to the scalar product and Case 2 to the scalar-tensor product (Def. \ref{def:stprod}; discussed in Sec. \ref{subsec:stprod}). In Case 3, $\mathcal{A}$ is always treated as a matrix and the \codebf{mvmul} command is simply generated (this accounts for Def. \ref{def:tinner} if $\mathcal{A}$ is a vector). 

From this point on, we define a function $f_n$ that refers to Case $n$ (e.g., $f_3$ generates an MVM instruction). Continuing, in Case 4, $\mathcal{A}$ is also treated as a matrix. $\mathcal{B}$ is decomposed into a chain of vectors and a series of references to Case 3 ($f_3(\mathcal{A}, \mat{b}_{i_1i_2 \dots i_{n-1}})$) are made. In Case 5, $\mathcal{A}$ is decomposed into a chain of matrices and a series of references to Case 3 ($f_3(\mat{A}_{i_1i_2 \dots i_{n-2}}, \mathcal{B})$) are again made.

In cases 6 and 7, we consider Def. \ref{def:tdot}. In Case 6, the tensor of lower rank is first decomposed. Case 4 is then referenced for each matrix if $A$ was decomposed (always into a matrix chain, resulting in calls to $f_4(\mat{A}_{i_1i_2 \dots i_{n-2}}, \mathcal{B})$) and Case 5 is referenced if $\mathcal{B}$ was decomposed (always into a vector chain, resulting in calls to $f_{5}(\mathcal{A}, \mat{b}_{i_1i_2 \dots i_{n-1}})$). Finally, in Case 7, $\mathcal{A}$ is decomposed and Case 4 is referenced ($f_{4}(\mat{A}_{i_1i_2 \dots i_{n-2}}$, $\mathcal{B})$).

\subsection{Vector Cross Product}
\label{subsec:cross}

The cross product is an operation that appears frequently in computational geometry/computer graphics. A common task is to generate a third vector orthogonal to two other vectors (or a plane formed by 3 points) \cite{eisele}. The cross product can also be used to calculate the distance between two lines and calculate if they are parallel. It also appears in a multitude of physics simulations.

For most applications, cross products are in $\mathbb{R}^3$ and between two vectors\footnote{Higher rank cross products can be defined using the Levi-Civita symbol $\epsilon_{ijk}$, which we omit due to relatively few applications.}. We consider the cross product in a positively oriented orthonormal basis. The cross product of two vectors in $\mathbb{R}^3$ as defined in Def. \ref{def:cross} is also given by the antisymmetric matrix-vector product

\[
\mat{a} \times \mat{b} = 
\begin{bmatrix}
0 & -a_3 & a_2\\
a_3 & 0 & -a_1\\
-a_2 & a_1 & 0\\
\end{bmatrix}
\begin{bmatrix}
b_1\\
b_2\\
b_3
\end{bmatrix}
\]
The \codebf{mvmul} command can simply be generated from here.

\subsection{Other Tensor Products}

The Khatri-Rao product is useful in variances in statistics, multi-way models, linear matrix equations, and signal processing \cite{Sims1990-uu, Chambers1998-kn, Bro-undated-hh, Lev-Ari-undated-vg, Budampati2003-hg}. The face-splitting product is useful in convolutional layers in neural networks and digital signal processing in a digital antenna array \cite{dsp, face-splitting}.

The code generation method shown in Sec. \ref{subsec:kron} can be easily extended to support the Khatri-Rao and face-splitting products given in definitions \ref{def:krao} and \ref{def:fsplit}  respectively. An \codebf{mvmul} command can be generated for each index $i$ on the operands $\mat{a_i}$ and $\mat{b_i}$.

\subsection{Compilation of Expressions}

Chaining multiple operations into expressions is supported. The code generator traverses the AST with the tensor algebra operator precedence discussed in Sec. \ref{subsec:grammar}, and each code generation command is called sequentially as outlined in Sec. \ref{sec:decomp}. However, as a prototype, the Apollo compiler assumes the arguments are valid and performs no expression-related error handling.

\section{Discussion}
There are still additions that will need to be made to the Apollo language in order to fully optimize optical computations. Most importantly, we will need to use our tensor storage algorithm only for highly sparse tensors involved in intermediate computations; we currently implement it for all tensors. We plan to also extend Apollo to generate t1926 and host instructions, integrate t1926 instructions with Tachyon 1, and develop the methodology by which Tachyon 1 would interact with the host CPU. Neural network activation functions, such as ReLU, sigmoid, and softmax, are planned to be hard-wired into the microchip; we will extend the language to support neural networks when this occurs. We also plan to add more useful tensor algebra operations based on the foundation discussed in this paper, such as the Matricized Tensor Times Khatri-Rao Product (MTTKRP). These and other extensions would help Apollo become a more robust and efficient language.

Future research should explore tensor storage methods that will be able to more efficiently represent sparse tensors while still making them easy to index into. In order to optimize for speed, it will also be crucial to investigate how to best minimize required communication between Tachyon 1 and the host CPU, as converting between optical and electrical signals takes a significant amount of time. We plan to conduct this research ourselves, but at the same time encourage others to look into it as well.

In the future, we plan on extending the advances made in developing the Apollo language to build APIs for high-level languages (e.g., Python, Java, C++, etc.) so that they will be able to utilize Tachyon 1. This will allow users of conventional languages to be able to harness the speed of optical computing for applications such as physics simulations and ML. We specifically plan on building libraries able to integrate with the TensorFlow and PyTorch APIs so that users will be able to run ML models made with these APIs on Tachyon 1.

Our current design framework for Apollo leads the way for more powerful calculations to be performed faster on a new generation of hardware. With future advancements and optimizations, Apollo has the potential to impact numerous fields in engineering, computer science, and the natural sciences by allowing for significantly faster tensor algebra computations.

\section{Conclusion}
\label{sec:conc}

In this paper, we show how to perform tensor algebra computations on an optoelectronic microchip through Apollo, a domain specific language designed for this purpose. We then go over the language, compiler, and virtual machine designs. Next, we show a new way to store tensors that outperforms both conventional storage and CSF format from a memory viewpoint while still being compatible with our optical hardware. Finally, we go over the compilation of tensor algebra expressions into matrix-vector multiplications, which are native to our microchip. We illustrate how complex tensor algebra expressions can be run quickly and efficiently through our methods. Finally, we discuss the impact of our research, provide suggestions for future research avenues, and outline how we plan to extend the Apollo language.

\subsection*{Acknowledgements}

We thank Dhruv Anurag for Apollo-related discussion and testing. We thank Jagadeepram Maddipatla for creating test cases. We thank Dr. Jonathan Osborne for mathematical discussion and advice. We thank Mr. Emil Jurj for supporting this project. We thank Shihao Cao for support and useful discussion about the project’s future. Finally, we thank our families for extended support and patience.

\nocite{}
\bibliography{references}

\begin{thebibliography}{10}
\providecommand{\url}[1]{#1}
\csname url@samestyle\endcsname
\providecommand{\newblock}{\relax}
\providecommand{\bibinfo}[2]{#2}
\providecommand{\BIBentrySTDinterwordspacing}{\spaceskip=0pt\relax}
\providecommand{\BIBentryALTinterwordstretchfactor}{4}
\providecommand{\BIBentryALTinterwordspacing}{\spaceskip=\fontdimen2\font plus
\BIBentryALTinterwordstretchfactor\fontdimen3\font minus
  \fontdimen4\font\relax}
\providecommand{\BIBforeignlanguage}[2]{{%
\expandafter\ifx\csname l@#1\endcsname\relax
\typeout{** WARNING: IEEEtran.bst: No hyphenation pattern has been}%
\typeout{** loaded for the language `#1'. Using the pattern for}%
\typeout{** the default language instead.}%
\else
\language=\csname l@#1\endcsname
\fi
#2}}
\providecommand{\BIBdecl}{\relax}
\BIBdecl

\bibitem{comsol}
\BIBentryALTinterwordspacing
``Comsol\&nbsp;multiphysics® software - understand, predict, and optimize.''
  [Online]. Available: \url{https://www.comsol.com/comsol-multiphysics}
\BIBentrySTDinterwordspacing

\bibitem{ansys}
\BIBentryALTinterwordspacing
``Engineering simulation software | ansys products.'' [Online]. Available:
  \url{https://www.ansys.com/products}
\BIBentrySTDinterwordspacing

\bibitem{multiphysics-modeling-veryst-engineering}
\BIBentryALTinterwordspacing
``Multiphysics modeling.'' [Online]. Available:
  \url{https://www.veryst.com/services/simulation-analysis/multiphysics-modeling}
\BIBentrySTDinterwordspacing

\bibitem{MultiphysicsSimulation}
\BIBentryALTinterwordspacing
D.~E. Keyes, L.~C. McInnes, C.~Woodward, W.~Gropp, E.~Myra, M.~Pernice,
  J.~Bell, J.~Brown, A.~Clo, J.~Connors, E.~Constantinescu, D.~Estep, K.~Evans,
  C.~Farhat, A.~Hakim, G.~Hammond, G.~Hansen, J.~Hill, T.~Isaac, X.~Jiao,
  K.~Jordan, D.~Kaushik, E.~Kaxiras, A.~Koniges, K.~Lee, A.~Lott, Q.~Lu,
  J.~Magerlein, R.~Maxwell, M.~McCourt, M.~Mehl, R.~Pawlowski, A.~P. Randles,
  D.~Reynolds, B.~Rivière, U.~Rüde, T.~Scheibe, J.~Shadid, B.~Sheehan,
  M.~Shephard, A.~Siegel, B.~Smith, X.~Tang, C.~Wilson, and B.~Wohlmuth,
  ``Multiphysics simulations: Challenges and opportunities,'' \emph{The
  International Journal of High Performance Computing Applications}, vol.~27,
  no.~1, pp. 4--83, 2013. [Online]. Available:
  \url{https://doi.org/10.1177/1094342012468181}
\BIBentrySTDinterwordspacing

\bibitem{pmlr-v139-blalock21a}
\BIBentryALTinterwordspacing
D.~Blalock and J.~Guttag, ``Multiplying matrices without multiplying,'' in
  \emph{Proceedings of the 38th International Conference on Machine Learning},
  ser. Proceedings of Machine Learning Research, M.~Meila and T.~Zhang, Eds.,
  vol. 139.\hskip 1em plus 0.5em minus 0.4em\relax PMLR, 18--24 Jul 2021, pp.
  992--1004. [Online]. Available:
  \url{https://proceedings.mlr.press/v139/blalock21a.html}
\BIBentrySTDinterwordspacing

\bibitem{DBLP:journals/corr/abs-2101-07107}
\BIBentryALTinterwordspacing
A.~Briola, J.~D. Turiel, R.~Marcaccioli, and T.~Aste, ``Deep reinforcement
  learning for active high frequency trading,'' \emph{CoRR}, vol.
  abs/2101.07107, 2021. [Online]. Available:
  \url{https://arxiv.org/abs/2101.07107}
\BIBentrySTDinterwordspacing

\bibitem{parker:2016:meng-thesis}
\BIBentryALTinterwordspacing
P.~A. Tew, ``An investigation of sparse tensor formats for tensor libraries,''
  M.Eng. Thesis, Massachusetts Institute of Technology, Cambridge, MA, Jun
  2016. [Online]. Available:
  \url{http://tensor-compiler.org/files/tew-meng-thesis-sparse.pdf}
\BIBentrySTDinterwordspacing

\bibitem{NEURIPS2021-b0ab42fc}
\BIBentryALTinterwordspacing
H.~Xu, K.~Kostopoulou, A.~Dutta, X.~Li, A.~Ntoulas, and P.~Kalnis,
  ``Deepreduce: A sparse-tensor communication framework for federated deep
  learning,'' in \emph{Advances in Neural Information Processing Systems},
  M.~Ranzato, A.~Beygelzimer, Y.~Dauphin, P.~Liang, and J.~W. Vaughan, Eds.,
  vol.~34.\hskip 1em plus 0.5em minus 0.4em\relax Curran Associates, Inc.,
  2021, pp. 21\,150--21\,163. [Online]. Available:
  \url{https://tinyurl.com/3vrf7crn}
\BIBentrySTDinterwordspacing

\bibitem{kjolstad:2017:taco}
\BIBentryALTinterwordspacing
F.~Kjolstad, S.~Kamil, S.~Chou, D.~Lugato, and S.~Amarasinghe, ``The tensor
  algebra compiler,'' \emph{Proc. ACM Program. Lang.}, vol.~1, no. OOPSLA, pp.
  77:1--77:29, Oct. 2017. [Online]. Available:
  \url{http://doi.acm.org/10.1145/3133901}
\BIBentrySTDinterwordspacing

\bibitem{doi:10.1137/1.9780898719918.ch7}
\BIBentryALTinterwordspacing
D.~M. Dunlavy, T.~G. Kolda, and W.~P. Kegelmeyer, \emph{7. Multilinear Algebra
  for Analyzing Data with Multiple Linkages}, pp. 85--114. [Online]. Available:
  \url{https://epubs.siam.org/doi/abs/10.1137/1.9780898719918.ch7}
\BIBentrySTDinterwordspacing

\bibitem{https://doi.org/10.7298/5ksm-sm92}
\BIBentryALTinterwordspacing
{Srivastava, Nitish Kumar}, ``Design and generation of efficient hardware
  accelerators for sparse and dense tensor computations,'' 2020. [Online].
  Available: \url{https://hdl.handle.net/1813/70449}
\BIBentrySTDinterwordspacing

\bibitem{lehrer-2010}
\BIBentryALTinterwordspacing
J.~Lehrer, ``1,084 days: How toy story 3 was made,'' Jun 2010. [Online].
  Available: \url{https://www.wired.co.uk/article/how-toy-story-3-was-made}
\BIBentrySTDinterwordspacing

\bibitem{DBLP:journals/corr/abs-1911-12604}
\BIBentryALTinterwordspacing
P.~Peltzer, J.~Lotz, and U.~Naumann, ``Eigen-ad: Algorithmic differentiation of
  the eigen library,'' \emph{CoRR}, vol. abs/1911.12604, 2019. [Online].
  Available: \url{http://arxiv.org/abs/1911.12604}
\BIBentrySTDinterwordspacing

\bibitem{osti-1349514}
\BIBentryALTinterwordspacing
T.~Kola, B.~W. Bader, E.~N. Acar~Ataman, D.~Dunlavy, R.~Bassett, C.~J.
  Battaglino, T.~Plantenga, E.~Chi, S.~Hansen, and USDOE, ``Tensor toolbox for
  matlab v. 3.0,'' 3 2017. [Online]. Available:
  \url{https://www.osti.gov/biblio/1349514}
\BIBentrySTDinterwordspacing

\bibitem{SPLATT}
S.~Smith, N.~Ravindran, N.~D. Sidiropoulos, and G.~Karypis, ``Splatt: Efficient
  and parallel sparse tensor-matrix multiplication,'' in \emph{2015 IEEE
  International Parallel and Distributed Processing Symposium}, 2015, pp.
  61--70.

\bibitem{Cole:21}
\BIBentryALTinterwordspacing
C.~Cole, ``Optical and electrical programmable computing energy use
  comparison,'' \emph{Opt. Express}, vol.~29, no.~9, pp. 13\,153--13\,170, Apr
  2021. [Online]. Available:
  \url{http://opg.optica.org/oe/abstract.cfm?URI=oe-29-9-13153}
\BIBentrySTDinterwordspacing

\bibitem{DBLP:journals/corr/abs-2102-06365}
\BIBentryALTinterwordspacing
S.~Garg, J.~Lou, A.~Jain, and M.~A. Nahmias, ``Dynamic precision analog
  computing for neural networks,'' \emph{CoRR}, vol. abs/2102.06365, 2021.
  [Online]. Available: \url{https://arxiv.org/abs/2102.06365}
\BIBentrySTDinterwordspacing

\bibitem{nisan-schocken-2021}
N.~Nisan and S.~Schocken, \emph{The Elements of Computing Systems: Building a
  modern computer from first principles}.\hskip 1em plus 0.5em minus
  0.4em\relax The MIT Press, 2021.

\bibitem{corob-msft}
\BIBentryALTinterwordspacing
Corob-Msft, ``Arrays (c++).'' [Online]. Available:
  \url{https://docs.microsoft.com/en-us/cpp/cpp/arrays-cpp?view=msvc-170}
\BIBentrySTDinterwordspacing

\bibitem{java-array}
\BIBentryALTinterwordspacing
``Arrays.'' [Online]. Available:
  \url{https://docs.oracle.com/javase/tutorial/java/nutsandbolts/arrays.html}
\BIBentrySTDinterwordspacing

\bibitem{LOAN200085}
\BIBentryALTinterwordspacing
C.~F. Loan, ``The ubiquitous kronecker product,'' \emph{Journal of
  Computational and Applied Mathematics}, vol. 123, no.~1, pp. 85--100, 2000,
  numerical Analysis 2000. Vol. III: Linear Algebra. [Online]. Available:
  \url{https://www.sciencedirect.com/science/article/pii/S0377042700003939}
\BIBentrySTDinterwordspacing

\bibitem{DeepKNN}
\BIBentryALTinterwordspacing
A.~D. Jagtap, Y.~Shin, K.~Kawaguchi, and G.~E. Karniadakis, ``Deep kronecker
  neural networks: {A} general framework for neural networks with adaptive
  activation functions,'' \emph{CoRR}, vol. abs/2105.09513, 2021. [Online].
  Available: \url{https://arxiv.org/abs/2105.09513}
\BIBentrySTDinterwordspacing

\bibitem{DotProductInML}
\BIBentryALTinterwordspacing
S.~Rabanser, O.~Shchur, and S.~Günnemann, ``Introduction to tensor
  decompositions and their applications in machine learning,'' 2017. [Online].
  Available: \url{https://arxiv.org/abs/1711.10781}
\BIBentrySTDinterwordspacing

\bibitem{DotProductInQuantumPhysics}
\BIBentryALTinterwordspacing
G.~Dahl, J.~M. Leinaas, J.~Myrheim, and E.~Ovrum, ``A tensor product matrix
  approximation problem in quantum physics,'' \emph{Linear Algebra and its
  Applications}, vol. 420, no.~2, pp. 711--725, 2007. [Online]. Available:
  \url{https://www.sciencedirect.com/science/article/pii/S0024379506004149}
\BIBentrySTDinterwordspacing

\bibitem{eisele}
\BIBentryALTinterwordspacing
R.~Eisele, ``3d cross product.'' [Online]. Available:
  \url{https://www.xarg.org/book/linear-algebra/3d-cross-product/}
\BIBentrySTDinterwordspacing

\bibitem{Sims1990-uu}
C.~A. Sims, J.~H. Stock, and M.~W. Watson, ``Inference in linear time series
  models with some unit roots,'' \emph{Econometrica}, vol.~58, no.~1, p. 113,
  Jan. 1990.

\bibitem{Chambers1998-kn}
R.~L. Chambers, A.~H. Dorfman, and S.~Wang, ``\BIBforeignlanguage{en}{Limited
  information likelihood analysis of survey data},''
  \emph{\BIBforeignlanguage{en}{J. R. Stat. Soc. Series B Stat. Methodol.}},
  vol.~60, no.~2, pp. 397--411, 1998.

\bibitem{Bro-undated-hh}
R.~Bro, \emph{Multi-way Analysis in the Food Industry. Models}.\hskip 1em plus
  0.5em minus 0.4em\relax Algorithms and Applications.

\bibitem{Lev-Ari-undated-vg}
H.~Lev-Ari, \emph{Efficient solution of linear matrix equations with
  applications to multistatic}.

\bibitem{Budampati2003-hg}
R.~S. Budampati and N.~D. Sidiropoulos, ``{Khatri-Rao} space-time codes with
  maximum diversity gains over frequency-selective channels,'' in \emph{Sensor
  Array and Multichannel Signal Processing Workshop Proceedings, 2002}.\hskip
  1em plus 0.5em minus 0.4em\relax IEEE, 2003.

\bibitem{dsp}
V.~Slyusar, ``New matrix operations for dsp,'' 11 1999.

\bibitem{face-splitting}
\BIBentryALTinterwordspacing
D.~Ha, A.~M. Dai, and Q.~V. Le, ``Hypernetworks,'' \emph{CoRR}, vol.
  abs/1609.09106, 2016. [Online]. Available:
  \url{http://arxiv.org/abs/1609.09106}
\BIBentrySTDinterwordspacing

\end{thebibliography}
\bibliographystyle{IEEEtran}

\end{document}